\documentclass[grl]{agu2001}
 \usepackage{graphicx}
\authorrunninghead{STARLAB PREPRINT 04062004-01}

\titlerunninghead{Sea state monitoring using GNSS-R}

\authoraddr{G. Ruffini, Research Department, Starlab, C. de l'Observatori, s/n, 08035 Barcelona, Spain. (giulio.ruffini@starlab.es)}
\begin{document}

%
%

\title{Sea state  monitoring using coastal  GNSS-R}
%

%
%




\authors{F. Soulat, M. Caparrini, O. Germain, P. Lopez-Dekker,  M. Taani and G. Ruffini\altaffilmark{1}}

\altaffiltext{1}
{Starlab, C. de l'Observatori Fabra s/n, 08035 Barcelona, Spain, http://starlab.es}
%
%
%



%
%


\begin{abstract}
We report on a coastal experiment to study  GPS L1 reflections. The campaign was  carried out at the Barcelona Port breaker and dedicated to the development of sea-state retrieval algorithms. 
 An experimental system built for this purpose collected and  processed GPS data to automatically generate a times series of the interferometric complex field (ICF). 
The ICF  was analyzed off line and compared to a simple developed model that relates ICF coherence time to the ratio of significant wave height (SWH) and mean wave period (MWP). The analysis using this model showed  good consistency between the ICF coherence time and nearby oceanographic buoy  data. Based on this result, preliminary conclusions are drawn on the potential of  coastal GNSS-R for sea state monitoring using semi-empirical modeling to relate GNSS-R ICF coherence time to SWH.
\end{abstract}

\begin{article}

\section{Introduction}

Two Global Navigation Satellite Systems (GNSS)  are presently operational: the US Global Positioning System (GPS), and, to some extent, the Russian GLObal NAvigation Satellite System (GLONASS). In the next few years, the European Satellite Navigation System (Galileo) will be deployed and GPS will be modernized, providing more frequencies and wider bandwidth civilian codes. When that happens, more than 50 GNSS satellites will be emitting  precise L-band spread spectrum signals which will remain available for at least a few decades. Although designed for localization, these signals will no doubt be used within the GCOS/GOOS (Global Climate Observing System/Global Ocean Observing System).  This paper addresses coastal applications of the emerging  field known as GNSS Reflections (GNSS-R),  a passive, all weather bistatic radar technology  exploiting reflected GNSS signals, which aims at providing instruments and techniques for remote sensing of the ocean surface (in particular, sea surface height and roughness) and the atmosphere over the oceans.

The  Oceanpal project at Starlab focuses on the development of technologies for operational in-situ or low-altitude water surface monitoring  using GNSS-R. Oceanpal (patent pending) is an offspring of technology developed within several ESA/ESTEC projects targeted on the exploitation of GNSS Reflections from space, following the proposal of  \markcite{{\it Mart\'\i n-Neira} [1993]}. This instrument is to provide low cost/low maintenance sea state and altimetry measurements for coastal applications with the inherent precision of GNSS technology.

We report here an experimental campaign dedicated to sea-state monitoring using coastal GPS Reflections  as gathered and processed by the Oceanpal instrument. Altimetry using GNSS-R is discussed elsewhere (e.g., in \markcite{{\it Caparrini et al.} [2003]} and \markcite{{\it Ruffini et al.} [2004]} and references therein).

Section~\ref{experiment} describes the experimental campaign (instrumentation and data records) and the ground truth used for  validation of the results. A geophysical analysis based on the correlation time of the reflected field is then discussed in Section~\ref{data_analysis}. The results are compared to wind speed and buoy observables---SWH and MWP.
\begin{figure}[b!]
 \centering
   \includegraphics[width=7.5cm]{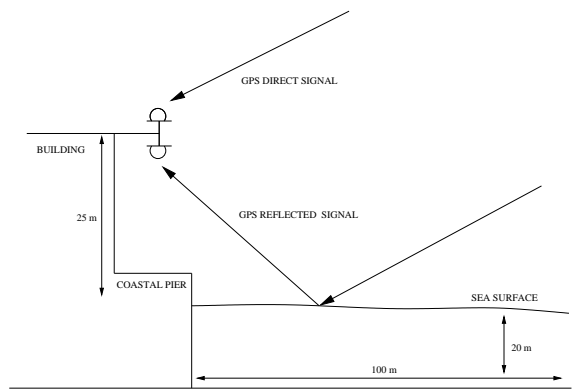}
   \caption{Sketch of the geometry of the coastal GNSS-R experiment.}
   \label{oceanpal_geometry}
\end{figure}

\section{Experimental campaign} 
\label{experiment}
\begin{figure}[b!]
   \includegraphics[width=7.5cm]{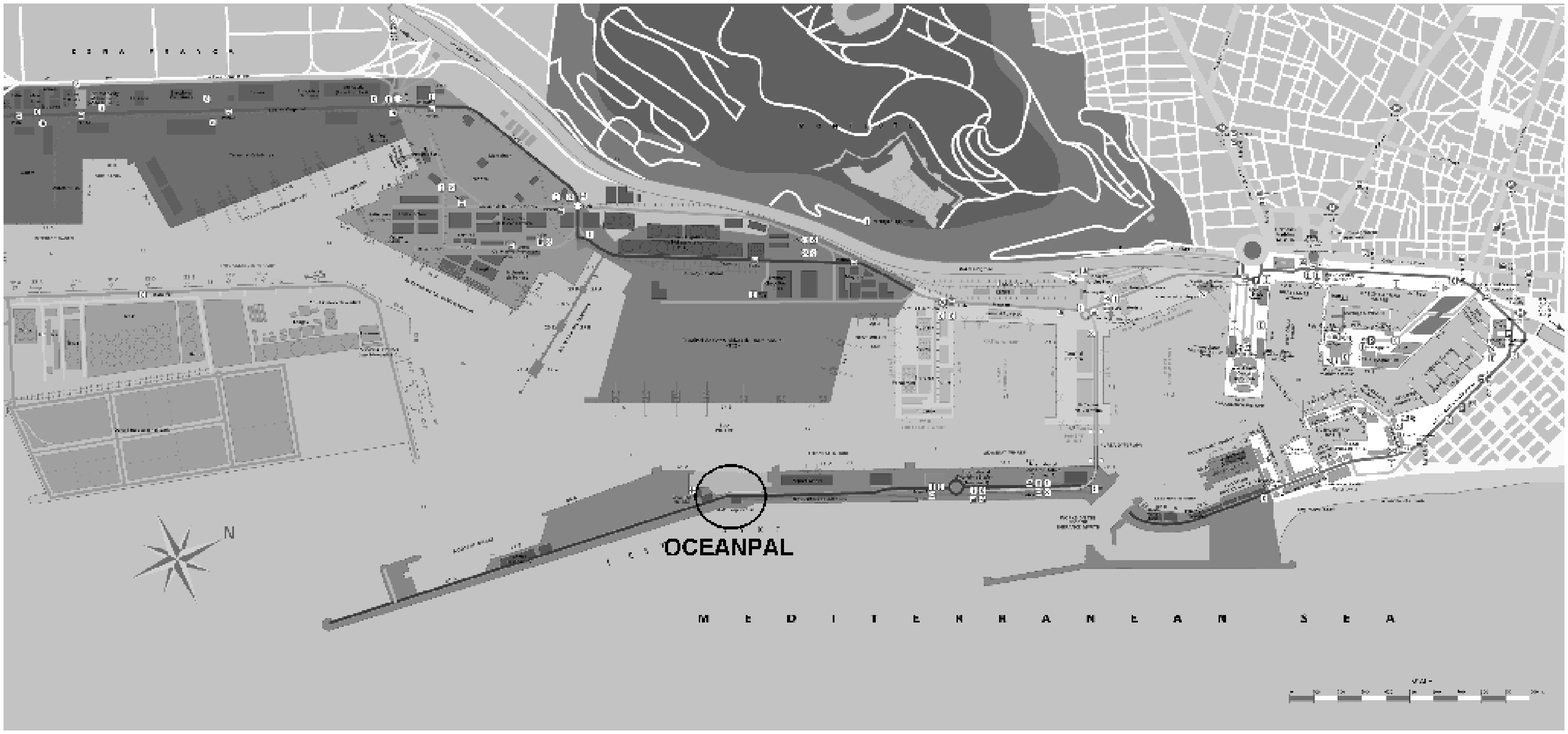}
   \caption{Map of the Barcelona Port area, and location of experiment (circle marked OCEANPAL)---courtesy of the Barcelona Port Authority.}
   \label{portmap}
\end{figure}

The HOPE 2 (Harbor Oceanpal Experiment) experimental campaign gathered  data from a ground platform in a range of sea-state conditions. It was carried out in December 2003 at the Meteorological station of Porta Coeli belonging to  the  Barcelona Port Authority, located on the port breakers (Figure \ref{portmap}). As sketched in Figure~\ref{oceanpal_geometry}, two antennas were deployed at approximately  25 meters  over the open sea surface to gather GPS-L1 signals. As usual in GNSS-R, one antenna was zenith-looking (Right Hand Circularly Polarized)  to get the direct signal, and the other was oriented towards the open sea surface with the complementary polarization (Left Hand Circularly Polarization) to get the reflected signals. The  antennas were connected to the instrument Front End, which then generated  a digital bit-stream of unprocessed GPS  data at IF. The IF data were recorded at a sample frequency of $\sim$16 MHz, processed on board to produce complex waveforms, and transferred  to the instrument  mass storage unit. The system also included a digital video camera providing  daytime optical records of  sea-state in the area (presence of swell, foam, breaking waves, calm waters...) during the periodic data acquisitions.

The campaign was divided into three parts (called Takes), each of them lasting 3 to 4 days. 
Take 1 was dedicated to the determination of the optimal position and orientation of the antennas given the experimental site and will not be described further.
Take 2 took place from December 5th until December 7th  2003, and Take 3 from the 12th to the 16th.

GPS-R and video data were gathered during 1 minute every 4 hours  in Take 2 and during 1 minute every 2 hours in Take 3, with otherwise exactly the same set-up. 

The instrument computed direct and reflected correlation waveforms using  a coherent integration time of 20 ms. Satellite visibility  was limited by the presence of the pier: the optimal mask angle was found to be [14$^o$-35$^o$] in elevation and [70$^o$-155$^o$] in azimuth. This mask excluded satellites for which  a clear impact of ground reflections on the complex reflected field (essentially due to the pier) could be observed, and represented an observation area of  $\sim$100 m radius into the open sea waters (with depths of about 20 m).

A meteorological station at the experimental site provided the wind vector every minute. In addition, a  Datawell ``Waverider'' buoy  located near the Llobregat Delta---10 miles South from the experimental site and  1.3 miles from the coast (sea depth of 45 m)---delivered hourly SWH and MWP. During the campaign, SWH ranged between 0.3 m and 1.2 m. 

\begin{figure}[t!]
 \centering
   \includegraphics[width=7cm]{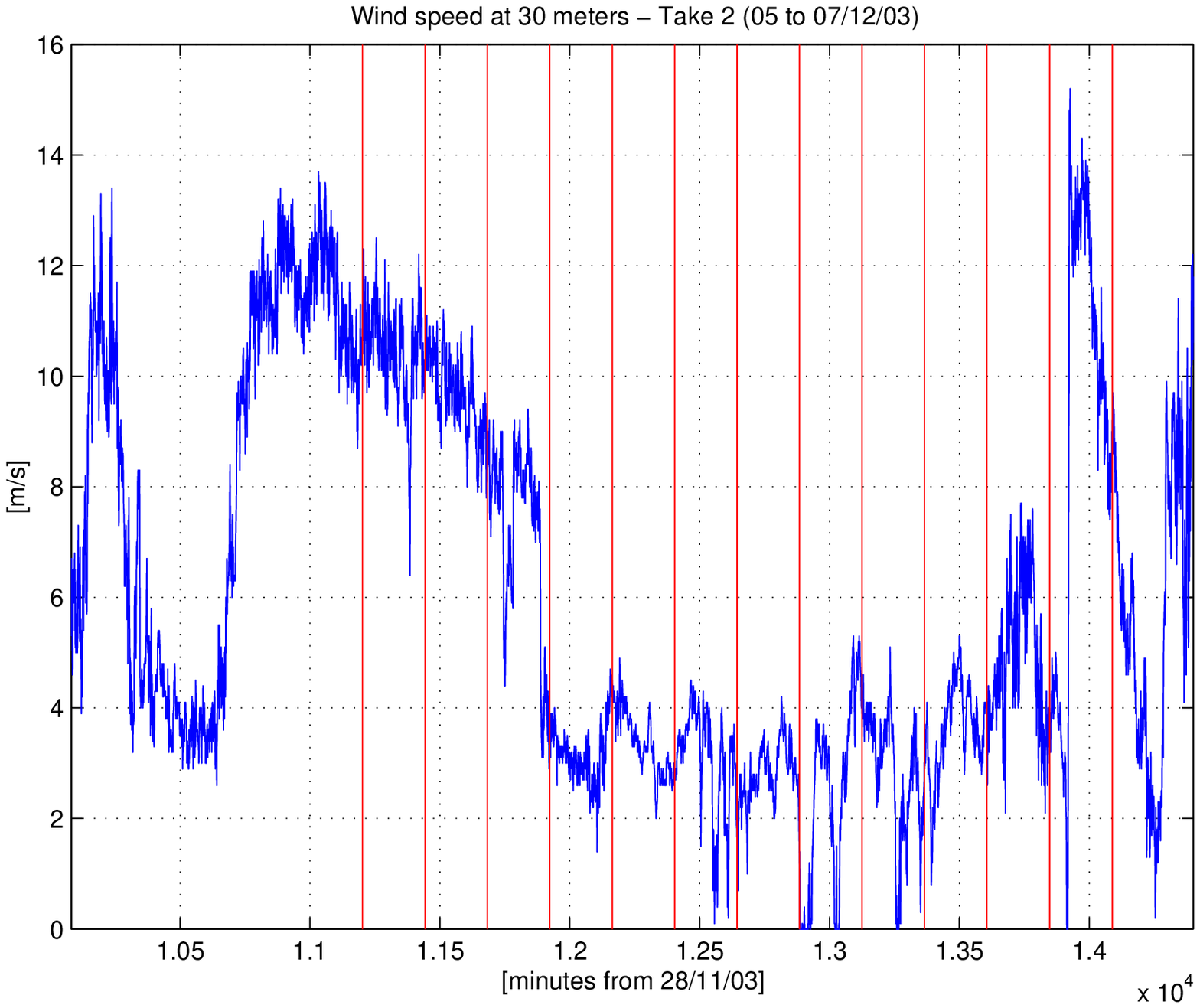} 
   \caption{Wind speed at the Barcelona Port Porta Coeli station during Take 2. Vertical bars  indicate the occurrence of Oceanpal GNSS-R measurements.}
   \label{take2}
\end{figure}

Figure~\ref{take2} shows wind speed during Take 2. The vertical bars indicate the occurrence of Oceanpal measurements. The wind speeds  observed during the data gathering ranged between 0 m/s and 12 m/s (U30). Observed wind directions indicated that during a significant portion  of the time the wind was blowing from land. Therefore,  the sea state was not expected to be significantly wind-related.

\section{Analysis}
\label{data_analysis}

After the down-converted and digitized GPS-L1 raw data  has been acquired, it is processed by a closed-loop GPS software receiver in the instrument data management unit (see \markcite{{\it Ruffini et al.} [2004]} for more details on the STARLIGHT tracking software). The software receiver initializes the process (that is, finds satellites in view and selects those within the mask), performs the correlations and tracks the delay and Doppler of the direct GPS-C/A signal, feeding this information to the reflected signal correlation module. This operation is typically carried out with 1--10 minute data segments (but limited to 1 minute in HOPE 2). The output of this process is so-called Level 0 data, consisting of the times series of complex waveforms for the direct and reflected signals. Level 0 data can then be used to produce Level 1 data (such as the ICF discussed below or the altimetric lapse discussed in \markcite{{\it Ruffini et al.} [2004]}) and  Level 2 geophysical products (such as sea surface height or sea state related parameters, as discussed next). The software receiver carries out other tasks, such as solving for the time and position of the up-looking antenna. 

The analysis for sea state begins with the interferometric complex field (ICF), defined at time $t$ 
by $F_I(t) = F_R(t)/F_D(t) $,
where $F_D$ and $F_R$ are the complex values at the amplitude peaks of the direct and reflected complex waveforms, respectively. 
The direct signal is thus  used  as a reference to remove features unrelated to ocean motion, such as any residual Doppler, the navigation bit phase offset, or direct signal power variability. The ICF contains very valuable information on the sea state. More precisely, it is the dynamics of the ICF which is of interest, as we discuss below.

We note that the coherent integration  process filters out high frequency components of the ICF (i.e.,  those $>$ 50 Hz for the used 20 ms). This is a factor to take into account for measurements in high seas (not the case in this experiment).

The goal of the analysis is to relate dynamics of the ICF to sea-surface geophysical parameters. Given the very small scattering area---we recall the instrument  was deployed at low altitude ($\sim 25$ m) in this experiment---sea-surface roughness parameters such as the Directional Mean Square Slope (DMSS) are not foreseen to be quantifiable  through the estimation of the width and orientation of the scattering area (Delay-Doppler mapping), especially given the coarse nature of the C/A code which is available today.

\subsection{ICF coherence time}
As a first step in the analysis of the ICF dynamics we have focused on the coherence time of the ICF, $\tau_{F}$,  defined here as the short time width of the ICF  autocorrelation function,  $ \Gamma(\Delta t) =  \langle F_I^*(t)F_I(t+\Delta t) \rangle_{z}$.

After removal of the carrier and code, we can use the  Fresnel integral approximation for the scattered field,   
\begin{equation}
F_R(t)= {- i  e^{ i  \Delta \omega t+in\pi} \over 4 \pi} \int \sqrt{G_{r}} {\cal R}\, { e^{ i  k(r+s)} \over r  s} \, (\vec{q} \cdot \hat{n})\,  dS ,
\end{equation}
where $\Delta \omega$ is the residual carrier frequency, $n\pi$ the navigation bit,  $\cal R$ is the Fresnel coefficient, $k=2\pi/\lambda$, with $\lambda \approx$19 cm in L1,  $r$ ($s$) is the distance between the receiver (transmitter) and each point of the sea-surface, $\hat{n}$ the surface normal  and  $\vec{q}=(\vec{q}_\bot, q_z)$ is  the scattering vector (the vector   normal to the plane that would specularly reflect the wave in the receiver direction). This vector is a function of the incoming and outgoing unit vectors $\hat{n}_i$ and $\hat{n}_s$, $\vec{q}= k(\hat{n}_i - \hat{n}_s)$.   
We assume here that $\vec{q} \cdot \hat{n}\approx k$ (small slope approximation, with scattering and/or support only near the specular).

We note here that the exponent in the integrand can be expanded as a power series in the surface elevation $z$, and that higher order terms are suppressed by the other scales in the problem. As an approximation, we can  retain only the first order term. In order to compute $ \Gamma(\Delta t)$ we now assume a Gaussian  probability distribution for the surface elevation and write (see, e.g., \markcite{{\it Beckmann and Spizzichino} [1963]})
\begin{equation}
\langle e^{-2 i  k\sin\epsilon[z(\vec{\rho},t)-z(\vec{\rho}',t+\Delta t)]}\rangle_{z} = e^{ -4k^2\sin^2\! \epsilon\, \sigma_{z}^2[1-C(\Delta \vec{\rho},\Delta t)]},
\end{equation}
where $ \vec{\rho}=(x,y)$ is the horizontal displacement vector from the specular point, $\sigma_{z}$ is the standard deviation of the surface elevation, $\epsilon$ the scattering elevation angle  and $C(\Delta \vec{\rho},\Delta t)$ the spatio-temporal autocorrelation function of the surface.
Using a parabolic isotropic approximation for $C(\Delta\vec{\rho},\Delta t)$ (valid for small $\Delta \vec{\rho}$ and $\Delta t$) and considering for simplicity that spatial and temporal properties of the surface can be separated, we  write $
C(\Delta \vec{\rho},\Delta t) \approx 1 - ( \Delta \rho)^2/2l_{z}^2 - {\Delta t^2}/2\tau_{z}^2$,
where $l_{z}$ and $\tau_{z}$ are, respectively, the correlation length and correlation time of the surface. Isotropy is a rather strong assumption, and will lead to a coherence time independent of wave direction (directional analysis will be taken up in a future effort).

\begin{figure*} [t!]
   \includegraphics[width=8.3cm]{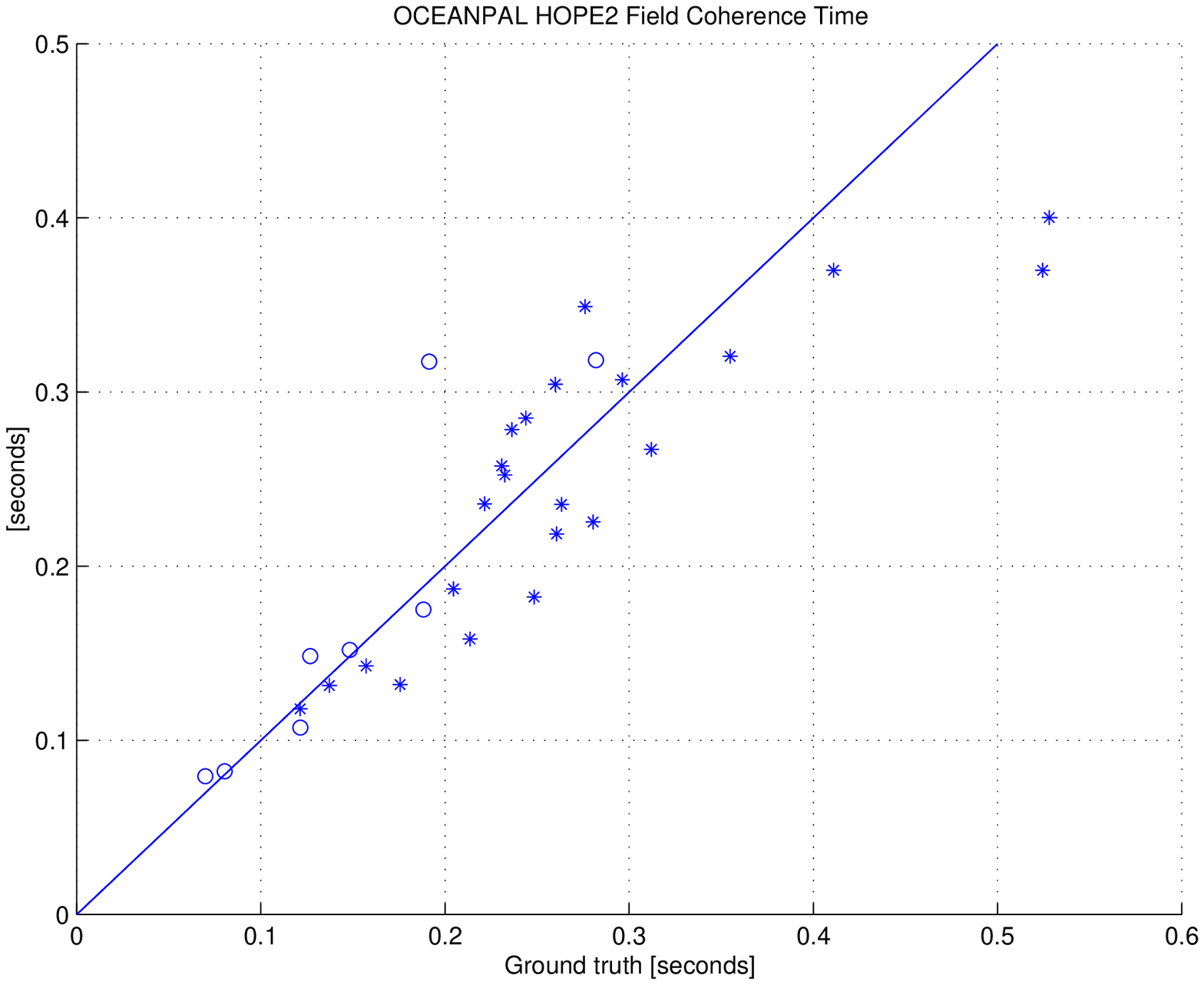}\ 
\includegraphics[width=8.3cm]{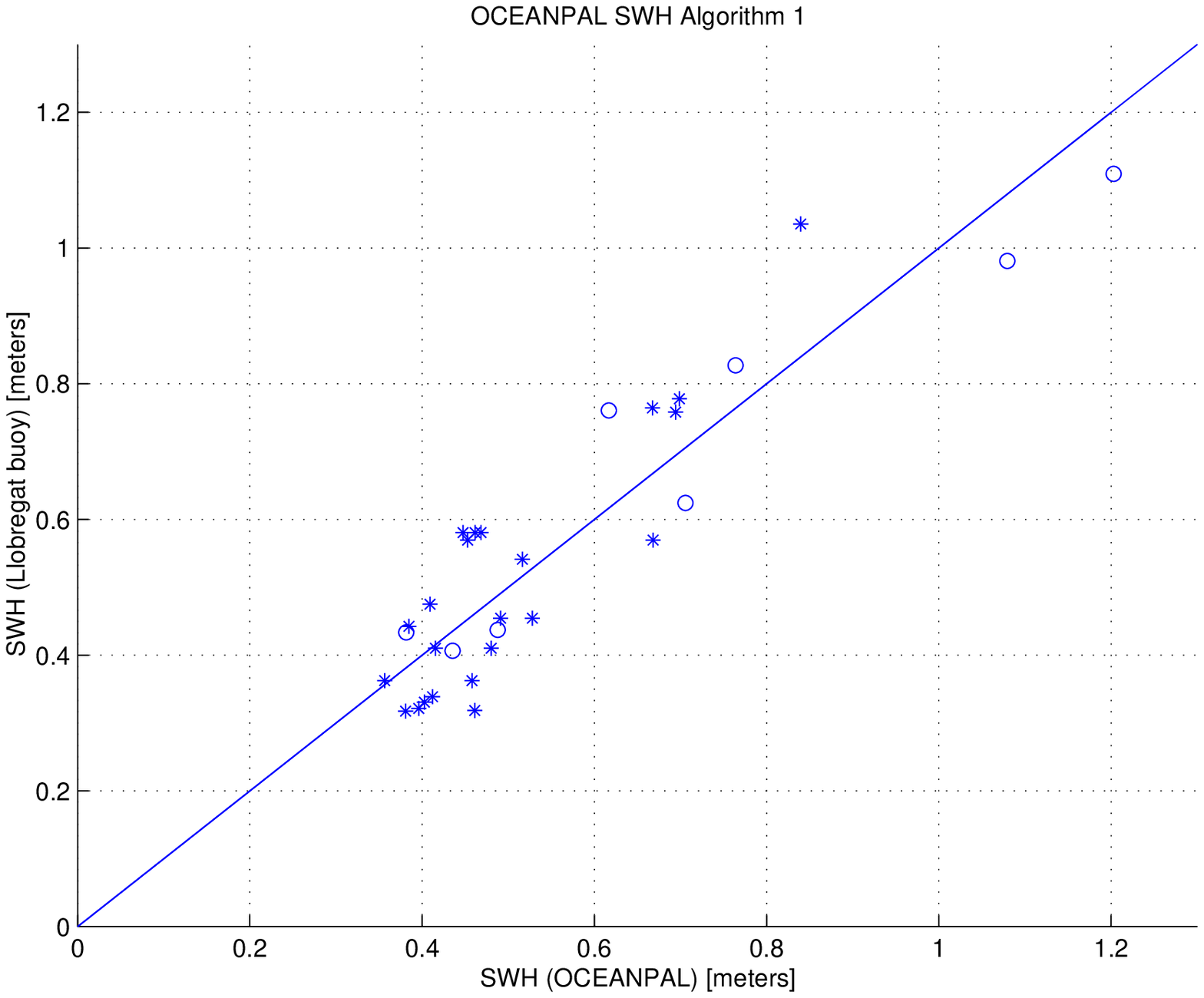}  
   \caption{Left: Measured ICF coherence time versus the estimate based on ground truth data (Equation~\ref{CT_field}). Right: Comparison of buoy SWH data with ICF coherence time SWH predictions using Oceanpal SWH Algorithm 1 described in Equation~\ref{oceanpalSWH}.  The algorithm standard deviation from the buoy data is 9 cm. Take 2 (circles) and Take 3 (stars) data are shown.}
   \label{comparison}
\end{figure*} 
Using this expression, it can be shown that the autocorrelation of the field can be approximated by
\begin{equation} \label{Corr_Func}
\Gamma(\Delta t) \approx A(\sigma_{z},l_{z},\epsilon,G_{r}) \, e^{-4k^2\sigma_{z}^2\frac{\Delta t^2}{2\tau_{z}^2}\sin^2\epsilon} .
\end{equation}
This equation, valid for small times,  states that the autocorrelation of the field is a Gaussian function of $\Delta t$ and proportional to a  coefficient depending on the sea surface elevation standard deviation $\sigma_{z}$, surface autocorrelation length, $l_{z}$,  geometry and  antenna gain $G_{r}$.

The coherence time of the ICF is now given by the width (second order moment) of this Gaussian function,
\begin{equation} \label{CT_field0}
\tau_{F} = \frac{ \tau_{z}}{2 k \sigma_{z} \sin \epsilon} = \frac{\lambda}{\pi \sin \epsilon}\frac{\tau_{z}}{\mbox{SWH}}.
\end{equation}
According to this model, $\tau_{F}$  depends on the electromagnetic wavelength  and the ratio between the correlation time of the surface and the significant wave height (an inverse velocity). A fundamental product of the instrument is therefore  $
\tau_z/\mbox{SWH} = \pi \tau_F \sin \epsilon /\lambda$.
 
In order to check this model using buoy data (SWH and MWP), we have derived a relation between MWP (available from the buoy measurements) and the sea-surface correlation time $\tau_{z}$ (needed to evaluate the right hand side of Equation~\ref{CT_field0}), through Monte-Carlo simulations using a Gaussian sea-surface spectrum (\markcite{{\it Elfouhaily} [1997]}). Simulating the surface propagation at a given point $z(x_o,y_o,t)$, the MWP was estimated through the Fourier analysis of the time series of $z(x_o,y_o,t)$, while $\tau_{z}$ was determined by the width of the autocorrelation function $\langle z(x_o,y_o,t)z(x_o,y_o,t+\Delta t)\rangle$. We obtained, for a well developed sea-state (with inverse wave age $\Omega=1$), the relation $ \tau_{z} \approx a_{m}+b_{m}\ast\mbox{MWP}$ ($a_{m}=0.07$, $b_m=0.12$, with an error of 0.09 s). Using this expression we can write
\begin{equation} \label{CT_field}
\tau_{F} \approx  \frac{\lambda}{\pi \sin \epsilon}\frac{ a_m+b_m\ast\mbox{MWP}  }{\mbox{SWH}}.
\end{equation}

Based on this  analysis, the Level 0 to Level 2 data processing involves two steps. First, the computation of the autocorrelation function of the complex interferometric field is carried out. Then,  a  
Gaussian
is fitted around lag zero 
to provide the estimate of the coherence time (Level 1).  

The comparison of the estimated ICF coherence time with the available ground truth  (wind speed, SWH and MWP) is made through Equation~\ref{CT_field}. The results are
shown in Figure~\ref{comparison} (left).  As observed, the measurements correlate well with theory.
Note that there is also  good consistency between Take 2 and Take 3 data.

It is worth mentioning that the linear  relationship  relating $ \tau_{z}$ to $\mbox{MWP}$ has been obtained under the assumption of a fully developed sea. This assumption will not hold in general in coastal areas for the whole range of sea-state conditions. 


\subsection{ICF and SWH}
Coherence time data  can  also  be translated into Level 2 geophysical products such as SWH using a semi-empirical algorithm, as we now  discuss.
We   assume that the correlation time of the surface is itself a function of the SWH and write an expression for the ``effective'' coherence time, $
\tau_F'\equiv \tau_{F} \sin\epsilon   = f({\mbox{SWH}})$,
where in the open ocean $f(\mbox{SWH})$ is in general a known function of SWH but which will also depend on the sea state maturity, fetch, bathymetry, etc. In coastal areas, this function will be harder to estimate from theory and a semi-empirical approach is envisioned. 

Based again on the \markcite{{\it Elfouhaily} [1997]} spectrum we have derived a linear relationship between $\tau_z$ and the SWH: $\tau_z=a_s+b_s\ast \mbox{SWH}$ ($a_s$=0.167, $b_s$=0.388, and an error of 0.03 s). This relation turns out to be rather independent of wave age. Using it, we can now rewrite Equation~\ref{CT_field0} as
\begin{equation} \label{CT_field2}
\tau'_{F} \approx \frac{\lambda}{\pi} \left( {a_s\over \mbox{SWH}} +b_s\right).
\end{equation}
Since the instrument gathered coastal data (within $\sim$100 m radius), the comparison with open ocean buoy data is not direct. 
In order to compare open ocean data  to coastal measurements, we include a SWH ``shift'' parameter, $\mbox{SWH}_0$ and a scale parameter $\gamma$. 
The algorithm for translation of effective ICF coherence time to SWH becomes, finally (Oceanpal SWH Algorithm 1),
\begin{equation} \label{oceanpalSWH}
\mbox{SWH}\approx \mbox{SWH}_0 + \gamma {a_s\over \tau'_{F} \pi/\lambda -b_s},
\end{equation}
valid for $\mbox{SWH}> \mbox{SWH}_0$.
We  have found that a value of $\mbox{SWH}_0$=0.21 m and $\gamma$=1.8 gives the best fit to the campaign data. Figure~\ref{comparison} (right) plots SWH buoy data against Oceanpal SWH Algorithm 1. The algorithm standard deviation from the buoy data is 9 cm.



\section{Conclusions}

GNSS-R is a budding new technology with a bright outlook. Scientific and operational  applications 
will clearly benefit from the precision, accuracy, abundance, stability and long-term availability of GNSS signals. 
The combination of GNSS-R data from air, ground and space can provide a coherent,  long lasting oceanographic monitoring infrastructure for decades to come.

In this paper we have highlighted an inexpensive, passive, dry operational sensor concept for use on coastal platforms and aircraft (for airborne applications focusing on sea state, see, e.g., \markcite{{\it Zavorotny et al.} [2000]} as well as  \markcite{{\it Germain et al.} [2003]}). Oceanpal is to provide precise sea level information and sea state, and we believe it will occupy an important niche in operational oceanography and marine operations. Other marine applications of this technology (salinity, pollution, currents) are also being studied.


The ground experiment described here showed that the coherence time of the interferometric field correlates well with the ratio of MWP and SWH, as predicted by the first order model summarized in Equation~\ref{CT_field}, and that a semi-empirical algorithm for SWH can also be devised.
As expected in a coastal area, no strong correlation was found between coherence time and wind speed, especially at low wind speed regimes and directions associated with land origin.  We also note that the buoy providing ground truth was located relatively far away from the experimental site (10 miles south) and in the open sea as opposed to close to the coast. Further experimental work with a closer buoy and/or on  an offshore platform will aid algorithm development. 

Based on this work, we foresee that ICF coherence time will play an important role in the production of Oceanpal sea state data, complementing well altimetric products. The derivation of a directional model and extraction of other information from the ICF time series are under development (\markcite{{\it Ruffini et al.} [2004b]}).

\setcounter{equation}{0}

%
%


%
%


%
%

\begin{acknowledgments}
We thank the Barcelona Port Authority Environmental Monitoring Department (APB) (J. Vil\'a, APB), the Polytechnic University of Catalunya/TSC (A. Camps) for experimental logistic support during the  campaigns and  to  the Catalan Meteorological Institute (SMC) for providing the buoy data. This work was partly supported by a Spanish Ministry of Science and Technology PROFIT project and the Starlab-IFREMER Salpex 2 project. We are also thankful for the support received in the context of several GNSS-R Starlab-ESA/ESTEC contracts dedicated to space applications, including  OPPSCAT 2 (3-10120/01/NL/SF),  Contract 15083/01/NL/MM (PARIS BETA),  Contract No. 14285/85/nl/pb, Starlab CCN3-WP3  and the ongoing  PARIS GAMMA project (TRP137).   
{All Starlab authors have contributed significantly; the Starlab author list has been ordered randomly.}
\end{acknowledgments}

\end{article}

\end{document}